\documentclass{kapproc} 

\usepackage{procps} 

\usepackage[dvips]{graphicx}

\upperandlowercase

\kluwerbib 

\begin{document}

\articletitle{UV Luminosity Function at $z$$\sim$4, 3, and 2}

\author{Marcin Sawicki}
\affil{Dominion Astrophysical Observatory, Herzberg Institute of Astrophysics, National Research Council,
5071 West Saanich Road, Victoria, BC, V9E 2E7, Canada}
\email{marcin.sawicki@nrc.ca}

\author{David Thompson}
\affil{Caltech Optical Observatories, California Institute of Technology,\\
MS 320-47, Pasadena, California, 91125, USA}
\email{djt@irastro.caltech.edu}

\begin{abstract}
We use very deep (${\cal R}_{lim}$=$27$) $U_nG{\cal R}I$ imaging to
study the evolution of the faint end of the UV-selected galaxy
luminosity function from $z$$\sim$4 to $z$$\sim$2.  We find that the
number of sub-$L^*$ galaxies increases from $z$$\sim$4 to $z$$\sim$3
while the number of bright ones appears to remain constant.  We find
no evidence for continued evolution to lower redshift, $z$$\sim$2.  If
real, this differential evolution of the luminosity function suggests
that \emph{differentially} comparing key diagnostics of dust, stellar
populations, etc.\ as a function of $z$ and $L$ may let us isolate the
key mechanisms that drive galaxy evolution at high redshift and we
describe several such studies currently underway.
\end{abstract}


\section{The Keck Deep Fields}

The shape of the galaxy luminosity function bears the imprint of
galaxy formation and evolutionary processes and suggests that galaxies
below $L^*$ differ substantially from those above it in more than just
luminosity. Our understanding of galaxy formation may profit from
studying the evolution of not just the bright but also the faint
component of the galaxy population at high redshift.

\begin{figure}
\includegraphics[width=\textwidth]{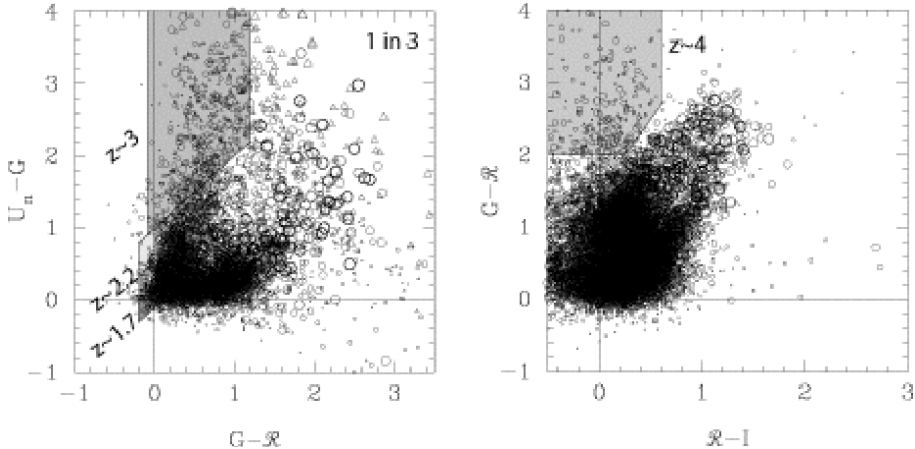} 
\label{colorcolor.fig}
\vspace{-5mm}
\caption{Color-color diagrams showing the selection regions for selecting 
high-$z$ samples in the KDF.  The left-hand panel illustrates
selection of $z$$sim$3, 2.2, and 1.7 samples; only a third of the
objects in our catalog are plotted.  The right hand panel shows the
selection of $z$$\sim$4 galaxies.  The filters we used and our color
selection are both {\emph {identical}} to those used by the Steidel
team to select galaxies at $z$$\sim$1.7--4.  However, our data reach
to ${\cal R}$=27, or 1.5 magnitudes deeper and thus significantly
below $L^*$ at these redshifts. {\emph {Contact authors for higher resolution Figure.}}}
\end{figure}

To study the evolution of the galaxy luminosity function at high
redshift, we have carried out a very deep imaging survey that uses the
very same $U_nG{\cal R}I$ filter set and color-color selection
technique as used in the work of Steidel et al.\ (1999, 2003, 2004),
but that reaches to ${\cal R}$=27 --- 1.5 magnitudes deeper and
significantly below $L^*$ at $z$=2--4 (Sawicki \& Thompson, 2005).
These Keck Deep Fields (KDF) were obtained with the LRIS imaging
spectrograph on Keck I and represent 71 hours of integration split
into five fields that are grouped into 3 spatially-independent patches
to allow us to monitor the effects of cosmic variance.  We use the
$U_nG{\cal R}I$ filter set and spec\-tros\-co\-pic\-al\-ly-confirmed
and -optimized color-color selection techniques developed by Steidel
et al.\ (1999, 2003, 2004).  Consequently, we can confidently select
sub-$L^*$ star-forming galaxies at $z$$\sim$4, 3, 2.2, and 1.7 without
the need for what at the magnitudes we probe would be extremely
expensive spectroscopic characterization of the sample.  To ${\cal
R}$=27, the KDF contains 427, 1481, 2417, and 2043, $U_nG{\cal
R}I$-selected star-forming galaxies at $z$$\sim$4, $z$$\sim$3,
$z$$\sim$2.2, and $z$$\sim$1.7, respectively, selected using the
Steidel et al.\ (1999, 2003, 2004) color-color selection criteria
(Fig.~1).  A detailed description of the Keck Deep Field observations,
data reductions, and the high-$z$ galaxy selection can be found in
Sawicki \& Thompson (2005).

\section{Luminosity function at high redshift}

\begin{figure}
\centerline{\includegraphics[height=7.5cm]{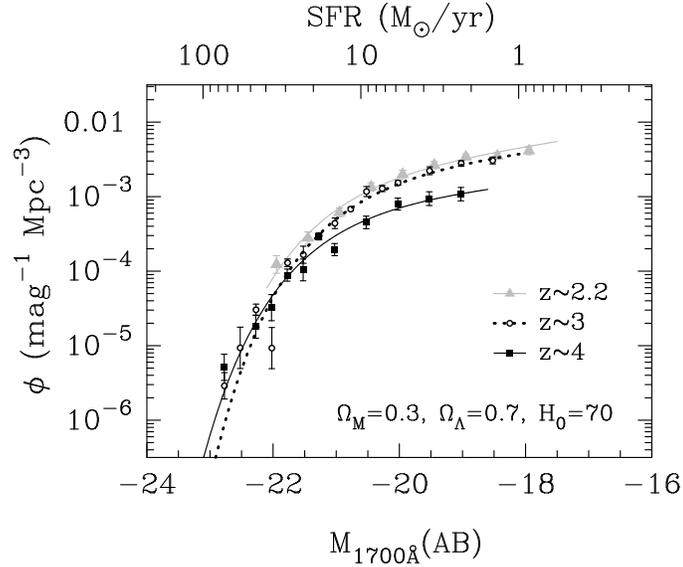} }
\label{LF.fig}
\vspace{0mm}
\caption{The luminosity function of high-$z$ rest-frame UV-selected galaxies
at $z$$\sim$4, $z$$\sim$3, and $z$$\sim$2.2.  No dust correction has
been applied; error bars include both $\sqrt{N}$ counting statistics
and field-to-field scatter.  There is a clear deficit of faint LBGs at
$z$$\sim$4 compared to $z$$\sim$3 and $z$$\sim$2.2. }
\end{figure}

Figure~2 shows the luminosity function of UV-selected star-forming
galaxies at $z$$\sim$4, 3, and 2.2.  At $z$$\sim$4 and 3, we augment
our KDF data with the identically-selected and similarly-computed
Steidel et al.\ (1999) LF measurements at bright magnitudes, ${\cal
R}$$<$25.5.  As in Steidel et al.\ (1999), our LF calculation uses the
effective volume technique, calculating $V_{eff}$ using recovery tests
of artificial high-$z$ galaxies implanted into the imaging data.  We
do not present the results for $z$$\sim$1.7 here because that analysis
is still ongoing as the narrow selection window in the $U_n$$-$$G$
color makes it necessary to carry out a more sophisticated treatment
of completeness and effective volume at $z$$\sim$1.7 than in the
higher redshift bins.  Our data reach down to very faint luminosities
which correspond to star formation rates (not adjusted for dust) of
only 1--2~$M_\odot$/yr (Fig.~2).

As Fig.~2 shows, we find a factor of $\sim$3 evolution in the number
counts (or, alternatively, 2 mag in luminosity) of \emph{faint} Lyman
Break Galaxies from $z$$\sim$4 to $z$$\sim$3 while at the same time,
there is no evidence for evolution from $z$$\sim$4 to $z$$\sim$3 at
the \emph{bright} end (Steidel et al.\ 1999; and Fig.~2).  Thus, it
seems that the luminosity function of high-$z$ galaxies evolves in a
\emph{differential} way, suggesting that different mechanisms drive
the evolution of the faint and of the luminous galaxies.  It is
unlikely that the observed evolution is due to a selection bias
because (1) if it were, we would expect the deficit to be present at
bright \emph{and} faint magnitudes, and (2) to make up the deficit
would require an unreasonably enormous expansion of the $z$$\sim$4
color selection box (Fig.~2).  We therefore conclude that the the
evolution from $z$$\sim$4 to $z$$\sim$3 is likely a reflection of a
true differential, luminosity-dependent evolutionary effect.  At the
same time, we see no evidence for evolution at the the faint end from
$z$$\sim$3 to $z$$\sim$2.2 (the bright end remains unconstrained at
present) suggesting that the mechanism responsible for the evolution
at earlier epochs saturates at lower redshifts.

\section{What is behind the evolution of the LF?}

At present, it is not clear what is responsible for the observed
differential evolution of the LF.  One straightforward possibility is
that the number of faint (but not bright) LBGs simply increases over
the 500~Myr from $z$$\sim$4 to $z$$\sim$3 (but not beyond).  Another
possibility is that dust properties --- its amount or covering
fraction --- may be decreasing in faint LBGs thus making them
brighter.  Alternatively, if star formation in individual faint LBGs
is time-variable, then they may brighten and fade (and thus move in
and out of a given magnitude bin in the LF) with duty cycle properties
that change with redshift.  However, whichever mechanism is
responsible, it appears to saturate by $z$$\sim$2.2.

The fact that the LF evolution appears to be differential suggests
that different evolutionary mechanisms are at play as a function of UV
luminosity. Studies that {\it differentially} compare key galaxy
properties as a function of luminosity and redshift should help us
isolate the mechanisms that are responsible for this evolution.  For
example: (1) The KDF is designed to measure galaxy clustering as a
function of both luminosity and redshift and doing so will let us
relate the potentially time-varying UV luminosity to the more stable
dark matter halo mass.  (2) We will also use a high-quality
$\sim$1000-hour 80-object \emph{composite} spectrum of a faint
$z$$\sim$3 LBG (Gemini GMOS observations are underway) to compare key
diagnostics of dust, superwinds, and stellar populations in a faint and
a luminous (e.g., Shapley et al.\ 2003) composite LBG. (3) Broadband
rest-frame UV-optical colors constrain the stellar population age and
the amount of dust in LBGs (e.g., Sawicki et al.\ 1998; Papovich et
al.\ 2001) and we can use this approach to look for systematic
differences in age and dust content.  Significantly, all such studies
will be making \emph{differential} comparisons, thereby reducing our
exposure to systematic biases in models or low-$z$ analogs.

A key point is that we have identified luminosity and redshift as
important variables in galaxy evolution at high $z$.  While LBG
follow-up studies to date have primarily focused on luminous galaxies
at $z$$\sim$3, extending such studies as a function of $z$ and $L$
should yield valuable insights into how galaxies form and evolve:
studying how key diagnostic properties of high-$z$ galaxies vary with
$L$ and $z$ will help us constrain what mechanisms drive galaxy
evolution in the early universe.


\end{document}